\documentclass[amssymb,aps]{revtex4}
\usepackage{graphicx}

\newcommand{\be}{\begin{equation}}
\newcommand{\ee}{\end{equation}}

\begin{document}

\title{The renormalization group improved effective potential in massless models}

\author{F. T. Brandt$^{a}$\footnote{fbrandt@usp.br},
F. A. Chishtie$^{b}$\footnote{fchishti@uwo.ca},
and D. G. C. McKeon$^{b}$\footnote{dgmckeo2@uwo.ca}}
\affiliation{$^a$ Instituto de F\'{\i}sica,
Universidade de S\~ao Paulo,
S\~ao Paulo, SP 05508-900, Brazil}

\affiliation{$^b$ Department of Applied Mathematics, The University of
Western Ontario, London, ON  N6A 5B7, Canada}

\bigskip

\begin{abstract}

The effective potential $V$ is considered in massless $\lambda\phi^4_4$
theory. The expansion of $V$ in powers of the coupling $\lambda$ and
of the logarithm of the background field $\phi$ is reorganized in two
ways; first as a series in $\lambda$ alone, then as a series in
$\ln\phi$ alone. By applying the renormalization group (RG) equation
to $V$, these expansions can be summed. Using the condition
$V^\prime(v) = 0$ (where $v$ is the vacuum expectation value of $\phi$)
in conjunction with the expansion of $V$ in powers of $\ln\phi$ fixes
$V$ provided $v\neq 0$. In this case, the dependence of $V$ on $\phi$
drops out and $V$ is not analytic in $\lambda$. Massless scalar
electrodynamics is considered using the same approach.

\end{abstract}

\maketitle

\section{Introduction}

The effective potential
\cite{Coleman:1973jx,Weinberg:1973ua,Jackiw:1974cv,Salam:1974rz}
has led to the suggestion that the renormalization induced scale
parameter $\mu$ might provide a scale parameter for a non vanishing
expectation value for a background scalar field $\phi$, even if there
is no mass parameter in the classical Lagrangian. This was realized 
%%%
{    in lowest order perturbation theory}
%%%
in the context of massless scalar electrodynamics (MSQED) \cite{Coleman:1973jx} 
and further examined in MSQED and
the standard model in ref. \cite{Elias:2003xp}.

The discussion of ref. \cite{Elias:2003xp} makes use of the RG equation
to sum leading log (LL) contributions to $V$. The condition
$V^\prime(v)=0$ then serves to relate gauge and scalar self couplings
arising in the models considered at a value of $\mu=v$. This was done
at one loop order in ref. \cite{Coleman:1973jx} in MSQED.

In this paper we first consider massless $\lambda\phi^4$. The RG
equation when applied to $V$ in this model can be used to determine in
closed form LL, NLL (next-to-leading log) etc, contribution to $V$,
much as has been done in
refs. \cite{Coleman:1973jx,Elias:2003xp}. However, it also proves to be
convenient to consider an alternative expansion of $V$ in powers of $\ln\phi$.
As was shown in the context of the effective action
\cite{Ahmady:2002qg}, the RG equation serves to fix all contributions to
$V$ dependent on $\ln\phi$ in terms of the contribution that is
independent of $\ln\phi$. When this is combined with the condition
$V^\prime(v)=0$, this $\ln\phi$-independent piece is fixed and $V$ is
determined completely provided $v\neq 0$. This is done either by
setting $\mu$ equal to $v$ or by treating $\mu$ as an independent
parameter. Remarkably, the resulting exact expansion for $V$ is not
analytic in $\lambda$ and is independent of $\phi$. If one were to avoid
these results, it is necessary to have $v=0$. (We do note though that
having $V$ independent of $\phi$ is consistent with the so called
``triviality'' \cite{Callaway:1988ya}.) Similar considerations are
applied to MSQED.
Once more it is found that the RG equation determines all $\ln\phi$
contributions to $V$ in terms of those that are independent of
$\ln\phi$. If $V^\prime(v)=0$ for all $\mu$, then again it follows
that $V$ is independent of $\phi$ provided $v\neq 0$.

\section{The RG equation and $V$ in $\lambda\phi^4$}

In the massless $\lambda\phi_4^4$ model, one can make the expansion
\begin{equation}\label{1}
V(\lambda,\phi,\mu) = \sum_{m=0}^{\infty}\sum_{n=0}^{m}\,T_{m,\,n}
\lambda^{m+1}\, L^{n}\,\phi^4
\end{equation}
where $L=\ln(\lambda\,\phi^2/\mu^2)$. (The dependence of $V$ on
$\ln\lambda$ arises when using some variant of the MS renormalization
scheme \cite{'tHooft:1973mm}; the counter term approach to
renormalization used in \cite{Coleman:1973jx} absorbs the dependence
on $\ln\lambda$ into 
%%%
{   a} 
%%%
finite renormalization.)

We first reorganize the double sum in Eq. (\ref{1}) so that
\begin{equation}\label{2}
V(\lambda,\phi,\mu) = \sum_{k=0}^{\infty}\,S_k(\lambda\,L)\,
\lambda^{k+1}\,\phi^4
\end{equation}
where
\begin{equation}\label{3}
S_k(x) = \sum_{l=0}^{\infty}\, T_{k+l,l}\,x^l .
\end{equation}
$S_0$ is the ``LL'' sum, $S_1$ is the ``NLL'' sum, etc. This sort of
reordering of the sum of Eq. (\ref{1}) has been used by Kastening
\cite{Kastening:1991gv} in conjunction with the effective potential,
as well as in a variety of other contexts in refs. 
\cite{Elias:2003xp,Ahmady:2002qg,Ahmady:2002fd,McKeon:2002yx,Elias:2002ry}.
(Other treatments of the effective potential appear in refs. 
\cite{Ford:1994dt,Chung:1999gi,Bando:1992wz,Casas:1994qy}.)

Renormalization induces dependence of $V$ on a scale parameter $\mu$
that does not occur in the initial Lagrangian. Explicit dependence on
$\mu$ is compensated for by implicit dependence on $\mu$ through the
renormalized couplings $\lambda(\mu)$ and the renormalized background
field $\phi(\mu)$; we have the RG equation
\begin{eqnarray}\label{4}
& \displaystyle{\mu\,\frac{d\;}{d\mu} V(\lambda(\mu),\phi(\mu),\mu) = 0} \nonumber\\
& =
\displaystyle{
\left(
\mu\frac{\partial}{\partial\mu} + \beta(\lambda(\mu))\,
\frac{\partial}{\partial\lambda(\mu)} + \gamma(\lambda(\mu))
\,\phi(\mu)\,
\frac{\partial}{\partial\phi(\mu)}
\right)\,V(\lambda(\mu),\phi(\mu),\mu) }
\end{eqnarray}
where
\begin{equation}\label{5}
\beta(\lambda(\mu)) = \mu\frac{d\lambda(\mu)}{d\mu} =
b_2\,\lambda^2+b_3\lambda^3+\cdots ,
\end{equation}
\begin{equation}\label{6}
\gamma(\lambda(\mu)) = \frac{\mu}{\phi(\mu)}
\frac{d\phi(\mu)}{d\mu} =
g_1\,\lambda+g_2\lambda^2+\cdots .
\end{equation}
In ref. \cite{McKeon:1994km} it is shown how the coefficients $b_i$,
$g_i$ can be determined by the coefficients $T_{m,n}$ in Eq. (\ref{1})
by using Eq. (\ref{4}). Here we wish to consider the inverse problem
of how portions of $V$ can be fixed from the RG equation (\ref{4}) if
$\beta$ and $\gamma$ are known. One approach would be to use the
method of characteristics 
as in \cite{Ford:1994dt,Chung:1999gi,Elias:2003cp,Elias:2004hi}. It is
also possible to substitute the expansions of Eqs. 
(\ref{2}), (\ref{5}) and (\ref{6}) into Eq. (\ref{4}) and equate
coefficients of $\lambda^k$ to zero. This results in a series of
coupled differential equations for $S_k(x)$ with the boundary
conditions 
\begin{equation}\label{7}
S_k(0) = T_{k,0}.
\end{equation}
(This approach has been used in 
refs. \cite{Elias:2003xp,Ahmady:2002qg,Kastening:1991gv,Ahmady:2002fd,McKeon:2002yx,Elias:2002ry}.)

The first two of these equations are
\begin{equation}\label{8}
(-2+b_2 x)S^\prime_0(x) + (b_2+4 g_1) S_0(x) = 0
\end{equation}
and
\begin{equation}\label{9}
(-2+b_2 x)S^\prime_1(x) + (2 b_2+4 g_1) S_1(x) 
+ (b_2 +2\, g_1 + b_3\, x)
S_0^\prime + (b_3+4g_2) S_0(x)= 0 .
\end{equation}
Eqs. (\ref{8}) and (\ref{9}) have solutions
\begin{equation}\label{10}
S_0(x) = T_{0,0} \left(1-\frac{b_2}{2}\, x\right)^{-1-4g_1/b_2}
\end{equation}
\begin{equation}\label{11}
S_1(x) = \frac{1}{w^2}\left[T_{1,0} -
\left(\frac{b_2}{2} + r\right)\ln(w) - s\,(w-1)\right]
\end{equation}
(In Eq. (\ref{11}), we have used the facts that $g_1 = 0$
and $T_{0,0} = 1$ and have set $r=b_3/b_2$, $s=4 g_2/b_2$
and $w=1-(b_2/2) x$.) These constitute the LL and NLL contributions
respectively to $V(\lambda(\mu),\phi(\mu),\mu)$ provided 
$x=\lambda(\mu)\ln[\lambda(\mu) \phi^2(\mu)/\mu^2]$.
They are analytic in $\lambda(\mu)$ about $\lambda(\mu)=0$
although there is a ``Landau pole'' when $w=0$.

Expanding about $\lambda(\mu)=0$ yields the one and two loop
contributions to $V$ coming from perturbation theory plus all RG
accessible portions of higher loop diagrams. Explicit calculation of
one-loop diagrams is needed to determine $T_{1,0}$. Subsequent
functions $S_n(x)$ which give N$^n$LL contributions to $V$ ($n>1$) can
be determined from Eq. (\ref{4}) provided $T_{n,0}$ is known.

There is an alternate to Eq. (\ref{2}) for reorganizing the double sum
in Eq. (\ref{1}) that has been used in
refs. \cite{Ahmady:2002qg,Elias:2004hi}. This is simply to expand in
powers of $L$ so that
\begin{equation}\label{12}
V(\lambda,\phi,\mu) = \sum_{n=0}^{\infty}\,A_n(\lambda)\, L^{n}\,\phi^4
\end{equation}
where
\begin{equation}\label{13}
A_n(\lambda) = \sum_{m=n}^{\infty}\, T_{m,\,n}\,\lambda^{m+1}.
\end{equation}
Substitution of Eq. (\ref{12}) into Eq. (\ref{4}) yields
\begin{equation}\label{14}
A_{n+1}(\lambda) = \frac{1}{n+1}\left[\frac 1 2 \,
\hat\beta(\lambda)\,A^\prime_n(\lambda) + 
2\,\hat\gamma(\lambda)\,A_n(\lambda) \right] 
\end{equation}
where
\begin{equation}\label{15}
\hat\beta  = \frac{\beta}{1-\gamma-\beta/2\lambda},\;\;\;
\hat\gamma = \frac{\gamma}{1-\gamma-\beta/2\lambda}.
\end{equation}
If $\beta$ and $\gamma$ are known, Eq. (\ref{14}) serves to fix $A_n$
($n>0$) in terms of $A_0$, and hence $V$ itself can be written in
terms of $A_0$. To do this, we follow the approach 
of refs. \cite{Ahmady:2002qg,Elias:2004hi}. This involves defining
\begin{equation}\label{16}
B_n(\lambda) = \exp\left(4\int_{\lambda_0}^\lambda
  \frac{\gamma(x)}{\beta(x)} \,dx\right)\, A_n(\lambda)
\end{equation}
and setting
\begin{equation}\label{17}
\eta(\lambda) = 
\int_{\lambda_0}^\lambda\left(\frac{2}{\hat\beta(x)}\right) dx 
\end{equation}
so that Eq. (\ref{14}) is re-expressed as
\begin{eqnarray}\label{18}
B_{n+1}(\lambda(\eta)) & = & \frac{1}{n+1}\frac{d}{d\eta}
B_n(\lambda(\eta)) \nonumber \\ 
& = & \frac{1}{(n+1)!}
\left(\frac{d}{d\eta}\right)^{n+1} \, B_0(\lambda(\eta))
\end{eqnarray}
Eq. (\ref{12}) then becomes, by Eqs. (\ref{16}) and (\ref{18})
\begin{eqnarray}\label{19}
V & = & \exp\left(-4\,\int_{\lambda_0}^\lambda
\frac{\gamma(x)}{\beta(x)} dx\right)\sum_{n=0}^\infty
\frac{L^n}{n!}\,
\left(\frac{d}{d\eta}\right)^{n} \, B_0(\lambda(\eta)) \, \phi^4
\nonumber \\
& = & \exp\left(-4\int_{\lambda_0}^\lambda\frac{\gamma(x)}{\beta(x)}\, dx\right)\,
B_0(\lambda(\eta+L)) \,\phi^4 \nonumber \\
& = & \exp\left(4\int_{\lambda}^{\lambda(\eta+L)}
\frac{\gamma(x)}{\beta(x)}\, dx\right)\,
A_0(\lambda(\eta+L)) \,\phi^4.
\end{eqnarray}
Now that $V$ has been expressed in terms of $A_0$, a second condition
can be imposed to fix $A_0$ itself. This condition is
\begin{equation}\label{20}
\frac{d}{d\phi}\, V(\lambda,\phi=v,\mu) = 0,
\end{equation}
where $v$ is an extremum of $V$. Eqs. (\ref{12}) and (\ref{20})
together lead to 
\begin{equation}\label{21}
\sum_{n=0}^{\infty} (4\, A_n(\lambda) + 2\, (n+1)\,
A_{n+1}(\lambda))\left(\ln\frac{\lambda\,v^2}{\mu^2}\right)^n\,v^3 = 0
\end{equation}
The parameter $\mu^2$ is not specified; $\lambda=\lambda(\mu)$ is a
function of $\mu^2$ whose evolution is dictated by Eq. (\ref{5}). In
refs. \cite{Coleman:1973jx,Elias:2003xp} this arbitrariness is
exploited in the context of MSQED by setting
$\mu^2=v^2$ in the equation $V^\prime(\phi=v)=0$ to obtain a relation
between the quartic scalar coupling $\lambda(\mu)$ and the gauge
coupling $e^2(\mu^2)$ evaluated at this value of $\mu^2$. In 
ref. \cite{Coleman:1973jx}, $V$ is just the one loop coupling, while
in ref. \cite{Elias:2003xp} the LL contributions to $V$ are
considered. Applying the condition of Eq. (\ref{20}) when $\mu^2=v^2$
when $V$ is the one loop does not give a consistent result, as can be
seen from the discussions of ref. \cite{Coleman:1973jx}. 
However, setting $\mu^2=\lambda(\mu)\,v^2$ 
so that all logarithms disappear in Eq. (\ref{21}) leads to
\begin{equation}\label{22}
A_1 = - 2\, A_0
\end{equation}
provided $v\neq 0$ as then only the $n=0$ term in Eq. (\ref{21})
survives. As the actual value of $\lambda(\mu^2=\lambda v^2)$ is an
independent quantity, dependent as it is on the constant of
integration arising in Eq. (\ref{5}), Eq. (\ref{22}) is a functional
equation. Together, Eq. (\ref{14}) with $n=0$ and Eq.(\ref{22}) show
that
\begin{equation}\label{23}
A_0^\prime(\lambda) + 
\left(\frac{4}{\beta(\lambda)} - \frac{2}{\lambda}\right)\, 
A_0(\lambda) = 0
\end{equation}
whose solution is
\begin{equation}\label{24}
A_0(\lambda) = K\,\exp\left[  
-\int_{\lambda_0}^\lambda\left(\frac{4}{\beta(x)} - \frac{2}{x} \right)
  \, dx\right]
\end{equation}
where $K = A_0(\lambda_0)$ is a boundary value. 
If $\beta(\lambda)$ is expanded as in Eq. (\ref{5}), then
Eq. (\ref{24}) can be rewritten as
\begin{equation}\label{25}
A_0(\lambda) = K^\prime\,\lambda^{(2+4b_3/b_2^2)}\,
\exp\left(\frac{4}{b_2\,\lambda}\right)\,
\exp\left\{-\int_0^\lambda\left[\frac{4}{\beta(x)}
- 4\left(\frac{1}{b_2\,x^2} - \frac{b_3}{b_2^2 x}\right)\right]dx\right\}.
\end{equation}
(The divergence appearing in the integral occurring in Eq. (\ref{24}) when
$\lambda_0\rightarrow 0$ has been absorbed into $K^\prime$.)
Together, Eqs. (\ref{19}) and (\ref{24}) show that
\begin{equation}\label{26}
V = K\exp\left(4\int_\lambda^{\lambda(\eta+L)}
\frac{\gamma(x)}{\beta(x)}\, dx\right)\exp
\left(-\int_{\lambda_0}^{\lambda(\eta+L)}\left(\frac{4}{\beta(x)} 
- \frac{2}{x}\right)dx\right) \,\phi^4,
\end{equation}
which becomes
\begin{equation}\label{27}
V = A_0(\lambda)\exp\left(\int_\lambda^{\lambda(\eta+L)}
\left(\frac{4\gamma(x) - 4 + 2\beta(x)/x}{\beta(x)} \right) dx\right)
\phi^4.
\end{equation}
From Eqs. (\ref{15}), (\ref{17}) and (\ref{27}) we obtain
\begin{eqnarray}\label{28}
V & = & A_0(\lambda)\exp\left\{-2\left[\eta\left(\lambda(\eta+L)\right)
-\eta\right]\right\}\,\phi^4 \nonumber\\
& = & A_0(\lambda)\,\exp(-2\,L)\,\phi^4.
\end{eqnarray}
The definition of $L$ reduces Eq. (\ref{28}) to
\begin{equation}\label{29}
V = \frac{A_0(\lambda)\,\mu^4}{\lambda^2},
\end{equation}
a result independent of $\phi$. It does however satisfy the RG
equation (\ref{4}).

%%%%
{   {
If a counter term renormalization were used to eliminate divergences
in the calculation of $V$, then the logarithm 
$L=\ln(\lambda\phi^2/\mu^2)$ appearing in Eq. (\ref{1}) would be
replaced by
\begin{equation}\label{30i}
\tilde L = \ln\left(\phi^2/\mu^2\right)
\end{equation}
as occurring in ref. \cite{Coleman:1973jx}. We would then have
\begin{equation}\label{31i}
V(\lambda,\phi,\mu)=\sum_{m=0}^{\infty}\sum_{n=0}^{m}
\tilde T_{m,n}\,\lambda^{m+1}\,\tilde L^{n} \,\phi^4
\end{equation}
with $T_{m,n}$ and $\tilde T_{m,n}$ related by a finite
renormalization dependent on $\ln\lambda$. If now
\begin{equation}\label{32i}
\tilde A_n(\lambda)= \sum_{m=n}^{\infty} \tilde T_{m,n}\,\lambda^{m+1}
\end{equation}
then Eq. (\ref{4}) results in
\begin{equation}\label{33i}
\tilde A_{n+1}(\lambda)=\frac{1}{n+1}\left[
\frac 1 2 \tilde\beta(\lambda)\tilde A^{\prime}_n(\lambda)+2
\tilde\gamma(\lambda)\tilde A_n(\lambda)\right] ,
\end{equation}
an equation similar to Eq. (\ref{14}), but now we have
\begin{equation}\label{34i}
\tilde\beta = \frac{\beta}{1-\gamma}, \;\;\;\;
\tilde\gamma = \frac{\gamma}{1-\gamma}
\end{equation}
in place of $\hat\beta$ and $\hat\gamma$ of Eq. (\ref{15}).

Just as Eq. (\ref{20}) leads to Eq. (\ref{22}), we find that
\begin{equation}\label{35i}
\tilde A_1 = - 2\, \tilde A_0 ;
\end{equation}
together Eqs. (\ref{33i}) and (\ref{35i}) lead to
\begin{equation}\label{36i}
\tilde A(\lambda)=\tilde K^{\prime}\,\lambda^{4 b_3/b_2^2}
\exp{\left(\frac{4}{b_2\lambda}\right)}
\exp{\left\{-\int_0^\lambda\left[\frac{4}{\beta(x)}-4
\left(\frac{1}{b_2 x^2}-\frac{b_3}{b_2^2 x}\right)\right] dx\right\}}
\end{equation}
showing that
\begin{equation}\label{37i}
\tilde A_0(\lambda) = \left(\tilde K^{\prime}/K^{\prime}\right)
A_0(\lambda)/\lambda^2
\end{equation}
with $A_0(\lambda)$ being given by Eq. (\ref{25}).

The steps leading to Eq. (\ref{19}) now result in
\begin{equation}\label{38i}
V=\exp{\left(4\int_\lambda^{\lambda(\tilde\eta+\tilde L)}
\frac{\gamma(x)}{\beta(x)} dx\right)}
\tilde A_0\left(\lambda(\tilde\eta + \tilde L)\right)\phi^4
\end{equation}
where in place of $\eta$ defined in Eq. (\ref{17})
\begin{equation}\label{39i}
\tilde\eta(\lambda) = 
\int_{\lambda_0}^\lambda\left(\frac{2}{\tilde\beta(x)}\right) dx ,
\end{equation}
so that in analogy to Eq. (\ref{28}) we arrive at
\begin{equation}\label{40i}
V=\tilde A_0\exp{\left(-2\tilde L\right)}\phi^4 =
\tilde A_0(\lambda)\, \mu^4.
\end{equation}
once again, $V$ is independent of $\phi$. From Eq. (\ref{37i}) we see
that the dependency of $V$ on $\lambda$ in Eqs. (\ref{29}) and
(\ref{40i}) is identical.

It is worth noting at this point that the renormalized coupling
defined in ref. \cite{Coleman:1973jx} to be 
$V^{\prime\prime\prime\prime}(\phi=\mu)$ 
is zero if $V$ is independent of $\phi$, as
is the case with $V$ given by Eq. (\ref{29}).}}
%%%%%%%

The only way to circumvent this conclusion that $V$ reduces to a
non-analytic function of $\lambda$ which is independent of $\phi$ is
to take the solution to Eq. (\ref{21}) to be
\begin{equation}\label{30}
v = 0
\end{equation}
rather than have Eq. (\ref{22}) satisfied. This implies that the symmetry
$\phi\leftrightarrow -\phi$ is unbroken by radiative corrections and
that $\phi$ remains massless.  The functions $A_n(\lambda)$ ($n>0$) are
still determined in terms of $A_0(\lambda)$ by Eq. (\ref{14}), but
$A_0(\lambda)$ is itself not fixed as Eq. (\ref{22}) cannot be invoked.

If Eq. (\ref{21}) were to hold for all $\mu$, then it is satisfied
order-by-order in $\ln(\lambda v^2/\mu^2)$ provided
\begin{equation}\label{31}
A_{n+1}(\lambda) = -\frac{2}{n+1}\, A_n(\lambda).
\end{equation}
Eq. (\ref{31}) reduces to Eq. (\ref{22}) when $n=0$; however now we
find that Eqs. (\ref{14}) and (\ref{31}) give rise to an equation for
all $A_n$, not just $A_0$. Eq. (\ref{24}) generalizes now to
\begin{equation}\label{32}
A_n(\lambda) = K_n\,\exp\left[
-\int_{\lambda_0}^\lambda\left(\frac{4}{\beta(x)} - \frac{2}{x} \right)
  \, dx
\right]
\end{equation}
with $K_{n+1} = -2 K_n/(n+1)$ in order for Eq. (\ref{31}) be
satisfied. Furthermore, from Eq. (\ref{31}) by itself,
\begin{equation}\label{33}
A_n(\lambda) = \frac{(-2)^n}{n!}\, A_0(\lambda)
\end{equation}
so that
\begin{equation}\label{34}
V = \sum_{n=0}^\infty \frac{(-2)^n}{n!}\, A_0(\lambda)
\left(\ln\frac{\lambda\phi^2}{\mu^2}\right)^n\,\phi^4
\end{equation}
which reproduces the result of Eq. (\ref{29}).

The approach of this section can be easily adapted to the effective
potential in a 
%%%
{   massless} 
%%%
$\lambda\phi^6_3$ model. In analogy with
Eq. (\ref{29}), we find that $V$ is independent of the background
field and non-analytic in the coupling $\lambda$ at $\lambda=0$.

\section{The RG equation and $V$ in MSQED}
The LL approximation to $V$ in MSQED has been considered in 
ref. \cite{Elias:2003xp}. Here we consider the consequence of making
an expansion of $V$ in MSQED that is similar to Eq. (\ref{12}),
\begin{equation}\label{35}
V(\lambda,e^2,\phi,\mu) = \sum_{n=0}^{\infty}\,A_n(\lambda,e^2)\, L^{n}\,\phi^4
\end{equation}
where again $L=\ln(\lambda\,\phi^2/\mu^2)$. The RG equation now becomes
\begin{equation}\label{36}
\left(
\mu\frac{\partial}{\partial\mu}+
\beta_\lambda(\lambda,e^2) \frac{\partial}{\partial\lambda}+
\beta_{e^2}(\lambda,e^2) \frac{\partial}{\partial e^2}+
\gamma(\lambda,e^2) \phi \frac{\partial}{\partial\phi}
\right)\, V = 0.
\end{equation}
Just as Eq. (\ref{14}) was derived, we find now from Eq. (\ref{36})
that
\begin{equation}\label{37}
A_{n+1} = \frac{1}{n+1}\left(
\frac 1 2 \, \hat\beta_\lambda \frac{\partial}{\partial\lambda}+ 
\frac 1 2 \, \hat\beta_{e^2} \frac{\partial}{\partial e^2}+ 
2\,\hat\gamma \right) A_n 
\end{equation}
where
\begin{equation}\label{38}
\hat\beta_\lambda  = \frac{\beta_\lambda}{1-\gamma-\beta_\lambda/2\lambda},\;\;
\hat\beta_{e^2}  = \frac{\beta_{e^2}}{1-\gamma-\beta_\lambda/2\lambda},\;\;
\hat\gamma = \frac{\gamma}{1-\gamma-\beta_\lambda/2\lambda}.
\end{equation}
From the analogue of Eq. (\ref{20}) 
\begin{equation}\label{39}
\frac{d}{d\phi}\, V(\lambda,e^2,\phi=v,\mu) = 0,
\end{equation}
we find that, much like Eq. (\ref{21}),
\begin{equation}\label{40}
\sum_{n=0}^{\infty} (4\, A_n(\lambda,e^2) + 2\, (n+1)\,
A_{n+1}(\lambda,e^2))\left(\ln\frac{\lambda\,v^2}{\mu^2}\right)^n\,v^3 = 0.
\end{equation}

If $v\neq 0$ and Eq. (\ref{40}) holds order-by-order in 
$\ln(\lambda v^2/\mu^2)$, then much like Eq. (\ref{31}) we have 
\begin{equation}\label{41}
A_{n+1}(\lambda,e^2) = -\frac{2}{n+1}\, A_n(\lambda,e^2)
\end{equation}
so that
\begin{equation}\label{42}
A_{n+1}(\lambda,e^2) = \frac{(-2)^{n+1}}{(n+1)!}\, A_0(\lambda,e^2).
\end{equation}
Together, Eqs (\ref{35}) and (\ref{42}) show that
\begin{equation}\label{43}
V(\lambda,e^2,\phi,\mu) = \frac{A_0(\lambda,e^2)\,\mu^4}{\lambda^2} ;
\end{equation}
Eq. (\ref{43}) is the generalization of eq. (\ref{29}) to
MSQED. Again, dependence on $\phi$ has disappeared and the dependence
on $\lambda$ is not analytic at $\lambda=0$. This situation is avoided
only if $v=0$ is taken to be the solution to Eq. (\ref{40}), in which
case there is not spontaneous symmetry breakdown. We note that even
with the flat effective potential of Eq. (\ref{43}), 
%%%
{   the model is not trivial and}
%%%
it is possible to
have $v\neq 0$ in which case the gauge field develops a mass.

In refs. \cite{Coleman:1973jx,Elias:2003xp} Eq. (\ref{39}) is only
applied when $\mu^2=\lambda(\mu) v^2$. At one loop 
(cf. ref. \cite{Coleman:1973jx}) or LL (cf. ref \cite{Elias:2003xp})
level, this results in a relationship between $\lambda$ and
$e^2$ at this mass scale.
More generally, from Eq. (\ref{39}) we find that when
$\mu^2=\lambda(\mu) v^2$
\begin{equation}\label{44}
A_1(\lambda,e^2) = - 2 A_0(\lambda,e^2)
\end{equation}
the $n=0$ limit of Eq. (\ref{41}), again provided $v\neq 0$. The
numerical values of $\lambda$ and $e^2$ can be independently varied as
they are contingent upon the boundary values for the functions
$\lambda(\mu)$, $e^2(\mu)$ and hence Eq. (\ref{44}) is a functional
relation for all $\lambda$, $e^2$.

We will now show that Eqs. (\ref{37}) and (\ref{44}) are sufficient to
establish Eq. (\ref{43}), much as Eqs. (\ref{14}) and (\ref{22}) are
shown to lead to Eq. (\ref{29}). Our method reduces to that outlined
in Eqs. (\ref{12}) to (\ref{29}) above in the limit $e^2=0$. It
involves adapting the ``method of characteristics'' 
(refs. \cite{Ford:1994dt,Chung:1999gi,Elias:2003cp,Elias:2004hi})
to the case in hand.

We begin by noting that from Eq. (\ref{37}) with $n=0$ and
Eq. (\ref{44}) that $A_0(\lambda,e^2)$ must satisfy
\begin{equation}\label{45}
\left[
\frac 1 2 \, \hat\beta_\lambda \frac{\partial}{\partial\lambda}+ 
\frac 1 2 \, \hat\beta_{e^2}   \frac{\partial}{\partial e^2}+ 
2\left(1+\hat\gamma \right)\right] 
A_0(\lambda,e^2) = 0.
\end{equation}
Characteristic functions $\bar\lambda(t)$ and $\bar e^2(t)$ are now
defined by
\begin{equation}\label{46}
\frac{d \bar\lambda(t)}{dt} = 
\frac 1 2\hat\beta_\lambda(\bar\lambda(t),\bar e^2(t)),\;\;
\bar\lambda(0)=\lambda
\end{equation}
\begin{equation}\label{47}
\frac{d \bar e^2(t)}{dt} = 
\frac 1 2\hat\beta_{e^2}(\bar e^2(t),\bar e^2(t)),\;\;
\bar e^2(0)= e^2
\end{equation}
We set
\begin{equation}\label{48}
a_n(\bar\lambda(t),\bar e^2(t),t) =
{\exp}\left\{ 2\int_0^t\hat\gamma(\bar\lambda(\tau),\bar e^2(\tau)) d\tau\right\}
A_n(\bar\lambda(t),\bar e^2(t))
\end{equation}
so that
\begin{eqnarray}\label{49}
\frac{d}{dt} a_n(\bar\lambda(t),\bar e^2(t),t) & = &
{\exp}\left\{2\int_0^t\hat\gamma(\bar\lambda(\tau),\bar e^2(\tau)) d\tau\right\}
\nonumber \\
& & \!\!\!\!\!\!\!\!\!\!\!\!\!\!
\left[
\frac 1 2 \, \hat\beta_\lambda(\bar\lambda,\bar e^2)\frac{\partial}
{\partial\bar\lambda}+ 
\frac 1 2 \, \hat\beta_{e^2}(\bar\lambda,\bar e^2)
\frac{\partial}{\partial\bar e^2}+
%\right. \nonumber \\ & & \left. \;\;\;\;\;\; 
2 \hat\gamma(\bar\lambda,\bar e^2)
\right] 
A_n(\bar\lambda,\bar e^2 ).
\end{eqnarray}
From Eqs. (\ref{37}), (\ref{48}) and (\ref{49}), we find that
\begin{equation}\label{50}
\frac{d}{dt} a_n(\bar\lambda(t),\bar e^2(t),t) = 
(n+1) a_{n+1}(\bar\lambda(t),\bar e^2(t),t) .
\end{equation}
Furthermore, we define
\begin{equation}\label{51}
V(t) = \sum_{n=0}^\infty a_n(\bar\lambda(t),\bar e^2(t),t)\bar L^n \phi^4
\end{equation}
where
\begin{equation}\label{52}
\frac{d\phi(t)}{dt} = \frac 1 2 \hat\gamma(\bar\lambda(t),\bar e^2(t))
\bar\phi(t),\;\; \bar\phi(0) = \phi
\end{equation}
and
\begin{equation}\label{53}
\frac{d\bar\mu(t)}{dt} = \frac 1 2
\frac{\bar\mu(t)}{1-\gamma(\bar\lambda(t), \bar e^2(t))
-\beta_\lambda(\bar\lambda(t), \bar e^2(t))/2\bar\lambda(t)}
,\;\; \bar\mu(0) = \mu
\end{equation}
with
\begin{equation}\label{54}
\bar L = \ln\left(\frac{\bar\lambda(t)\bar\phi^2(t)}{\bar\mu^2(t)}\right).
\end{equation}
By Eqs. (\ref{48}) and (\ref{49}), it is apparent that
\begin{equation}\label{55}
\frac{d V(t)}{dt} = 0
\end{equation}
and
\begin{equation}\label{56}
V(0) = V(\lambda,e^2,\phi,\mu^2).
\end{equation}
By Eq. (\ref{50}) we see that
\begin{equation}\label{57}
a_n(\bar\lambda(t), \bar e^2(t), t) = \frac{1}{n!}
\left(\frac{d}{dt}\right)^n a_0(\bar\lambda(t), \bar e^2(t), t)
\end{equation}
so that Eq. (\ref{51}) becomes
\begin{equation}\label{58}
V(t) = \sum_0^\infty\frac{\bar L^n}{n!} 
\left(\frac{d}{dt}\right)^n a_0(\bar\lambda(t), \bar e^2(t), t) \phi^4
\end{equation}
or
\begin{equation}\label{59}
V(t) = a_0(\bar\lambda(t+\bar L,\bar e^2(t+\bar L), t+\bar L) \phi^4
\end{equation}
If we set
\begin{equation}\label{60}
\tilde A_0(\bar\lambda(t),\bar e^2(t),t) =
{\exp}\left\{2\int_0^t\left[1+\hat\gamma(\bar\lambda(\tau),
\bar e^2(\tau),\tau)\right] 
d\tau\right\} A_0(\bar\lambda(t),\bar e^2(t))
\end{equation}
then by Eq. (\ref{45})
\begin{equation}\label{61}
\frac{d}{dt} \tilde A_0(\bar\lambda(t), \bar e^2(t),t) = 0
\end{equation}
with
\begin{equation}\label{62}
\tilde A_0(\bar\lambda(0), \bar e^2(0),0) = A_0(\lambda, e^2).
\end{equation}
Together, Eqs. (\ref{48}) and (\ref{60}) imply that
\begin{equation}\label{63}
a_0(\bar\lambda(t),\bar e^2(t), t) = {\exp}(-2t)
\tilde A_0(\bar\lambda(t), \bar e^2(t),t)
\end{equation}
so that Eq. (\ref{59}) becomes
\begin{eqnarray}\label{64}
V(t) & = & {\exp}\left(-2(t +\bar L)\right)\tilde 
A_0(\bar\lambda(t+\bar L),\bar e^2(t+\bar L), t+\bar L) \phi^4 \nonumber \\
& = & {\exp}(-2 t) 
\frac{
\tilde A_0(\bar\lambda(t + \bar L), \bar e^2(t + \bar L), t + \bar L) 
\bar\mu^4(t) \phi^4}
{\bar\lambda^2(t) \bar\phi^4(t)} .
\end{eqnarray}
From Eqs. (\ref{55}), (\ref{56}) and (\ref{64}) we obtain
\begin{equation}\label{65}
V(\lambda,e^2,\phi,\mu) = 
\frac{\tilde A_0(\bar\lambda(L), \bar e^2(L), L) \mu^4} 
{\lambda^2} ,
\end{equation}
which when combined with Eqs. (\ref{61}) and (\ref{62}) gives
\begin{equation}\label{66}
V(\lambda,e^2,\phi,\mu) = \frac{A_0(\lambda,e^2) \mu^4}{\lambda^2} .
\end{equation}
Eq. (\ref{43}) is consequently recovered using Eqs. (\ref{37}) and
(\ref{44}) alone.

\section{Discussion}
We have considered the effective potential $V$, first in the massless
$\lambda\phi^4$ model. The RG equation has been used to generate a
closed form expression for all LL contributions to $V$ from the one
loop result; in general all N$^n$LL contributions can be found
provided the $n$-loop contribution to $V$ has been computed.

In addition, it has been possible to express all log dependent
contributions to $V$ in terms of the log independent piece. When the
RG equation is supplemented by $V^\prime(\phi=v)=0$, then if $v\neq 0$,
this log independent piece of $V$ has been calculated, showing $V$ to
be independent of $\phi$ and non-analytic in $\lambda$ at $\lambda=0$.

These results are then shown to also hold in the MSQED model. It is
likely that the arguments apply in arbitrary massless models in which
scalars couple to gauge Bosons and Fermions.

%%%%
{   By having included all log dependent contributions to $V$ in these two
models and summing them through application of the RG equation and the
condition $V^{\prime}(\phi=v)=0$, we have come to a result at odds
with the scenario envisioned in ref. \cite{Coleman:1973jx} for
massless models involving scalars; $V(\phi)$ is now flat, having
neither a local minimum nor a singularity associated with a ``Landau pole''.
The discrepancy arises as a result of the way the condition of 
Eq. (\ref{20}) has been used. For us it implies a relationship between
$A_1$ and $A_0$ (given by Eqs. (\ref{22}) and (\ref{44})),
where $A_0$ and $A_1$ are respectively the contributions to $V$ that
are log independent and leading log to all orders. In ref. 
\cite{Coleman:1973jx}, these conditions are applied to the lowest
order perturbative contribution to $V$. This implies in the massless
$\phi^4_4$ model that the minimum $v$ lies outside the perturbative
range in the expansion for $V$ as $\lambda \ln(v^2/\mu^2)$ is
exceedingly large. For MSQED it is used to imply a relationship
between the quartic scalar coupling and the electric charge, evaluated
at a mass scale equal to the expectation value of $v$. This is a one
loop limit of Eq. (\ref{44}). By using the all orders (rather than the
one loop order) contributions $A_0(\lambda,e^2)$ and 
$A_1(\lambda,e^2)$ we are able to relate these two functions at this
mass scale; rather than arguing that this fixes $\lambda$ in terms of
$e^2$, we take these couplings to have independent values and the
functions $A_0$ and $A_1$ to be dependent on each other according to
Eq. (\ref{44}), allowing us to eventually demonstrate that $V$ is
independent of $\phi$ if $v\neq 0$.}
%%%

The effect on this analysis of including mass parameters into the
initial classical Lagrangian is clearly an outstanding problem that
needs to be addressed.

\section*{Acknowledgments}

We would like to thank A. Buchel, V. Elias, E. V. Gorbar, R. B. Mann,
V. A. Miransky, A. Rebhan, T. N. Sherry and T. G. Steele for helpful discussion.
NSERC and Fapesp provided financial support. 
DGCM would like to thank 
Perimeter Institute for Theoretical Physics for its
hospitality where part of this work was done and FTB would like
to thank the Department of Applied Mathematics of The University of
Western Ontario for the its hospitality.
R. and D. MacKenzie had a useful suggestion.

\end{document}